\documentstyle[preprint,aps]{revtex}

\begin{document}

\hsize = 6.5in
\widetext
\draft
\tighten
\topmargin-48pt
\evensidemargin5mm
\oddsidemargin5mm

\preprint{EFUAZ FT-96-40-REV}

\title{ More about the $S=1$ relativistic  oscillator\thanks{Submitted to
``Revista Mexicana de F\'{\i}sica".}}

\author{{\bf Valeri V. Dvoeglazov}\thanks{On leave of absence from
{\it Dept. Theor. \& Nucl. Phys., Saratov State University,
Astrakhanskaya ul., 83, Saratov\, RUSSIA.} Internet
address: dvoeglazov@main1.jinr.dubna.su}}

\address{
Escuela de F\'{\i}sica, Universidad Aut\'onoma de Zacatecas \\
Apartado Postal C-580, Zacatecas 98068, ZAC., M\'exico\\
Internet address:  valeri@cantera.reduaz.mx\\
URL: http://cantera.reduaz.mx/\~~valeri/valeri.htm}

\date{First version: December 1996, Final version: November 1997}

\maketitle

\begin{abstract}
Following to the lines drawn in my previous paper about the $S=0$
relativistic oscillator  I build up an oscillatorlike system  which can be
named as the $S=1$ Proca oscillator.  The Proca field function is
obtained in the framework of the Bargmann-Wigner prescription and
the interaction is introduced similarly to the $S=1/2$ Dirac oscillator
case regarded by Moshinsky and Szczepaniak. We obtained the intriguing
rule of quantization: ${\cal E} = \hbar \omega /2$ for the parity states
$(-1)^j$ and ${\cal E} = \pm \hbar \omega (j+1/2)$ for the parity states
$-(-1)^j$. There are no radial excitations. Finally, I apply the
above-mentioned procedure to the case of the two-body relativistic
oscillator.
\end{abstract}


\newpage

\baselineskip13pt

\section{The Proca Oscillator --- Puzzled Quantization}

In this Letter I continue  the study of oscillatorlike
systems first undertaken  by M. Moshinsky and A. Szczepaniak,
Ref.~\cite{Mosh}, for the $S=1/2$ case. Extensions of this model
to the case of a two-body problem and to the cases
of other spins as well have been presented in Ref.~\cite{Mosh2} and in
Refs.~\cite{Bruce,Deb,Ned}, respectively.
Moreover, a detailed consideration of the $S=1/2$ case,
which was presented in Ref.~\cite{oth},  demonstrated that
this form of interaction is free of the problem known as
the Klein paradox.  From the formal point of view the oscillatorlike
interaction can be interpreted as the interaction with a linear electric
field $E^i= \kappa r^i$, ref.~\cite{Moreno}, through the term of
$\sigma_{\mu\nu}F^{\mu\nu}/2$.

In my previous works, Refs.~\cite{DV0,DV1,DV2,DV3},
several interesting features of oscillatorlike systems
have been found. For instance, in Refs.~\cite{DV0,DV2}
the possibility of oscillatorlike construct
for arbitrary spin in the Dowker's formalism~\cite{Dowker}
has been proved. Some ideas providing a basis for
the matrix construct of the Klein-Gordon oscillator~\cite{Bruce}
have been presented in Refs.~\cite{DV0,DV1}.
In Ref.~\cite{DV3} the Bargmann-Wigner (BW) set of
equations is considered, with an {\it antisymmetric} second-rank spinor
being chosen as a field function. Such a description led to the
Kemmer-Dirac formalism~\cite{Kemmer} for the $S=0$ particle.  The
interaction introduced  in each of the BW equations
in the form proposed by Moshinsky and Szczepaniak
results in the oscillatorlike equation with the
double degeneracy (in $N$, the principal quantum number) of the spectrum
in the low frequency limit ({\it cf.} with Ref.~\cite{Bruce}).

In my opinion, all the above-said encourages further investigations.
The aim of this paper is to consider the $S=1$ Proca oscillator\footnote{I
take a liberty to name the equations obtained below as the Proca
oscillator since in a free case the equations (\ref{P1}-\ref{P4})  are the
well-known Proca equations, Ref.~\cite{Proca}.} with the interaction
introduced in the same manner as in ref.~\cite{DV3}. Namely, I
start from the Bargmann-Wigner equations with the non-gauge interaction
obtained after the substitution $\partial_i \rightarrow \partial_i
+ik\gamma^0 r^i$,\, $i=1,2,3$
\begin{eqnarray}\label{eq:BW1}
\cases{\left [
i\gamma^\mu \partial_\mu -k \gamma^i \gamma^0 r^i - m \right ] \Psi (x) =
0  &\cr &\cr \Psi (x) \left [ i (\gamma^\mu)^{^T} \partial_\mu - k
(\gamma^i \gamma^0)^{^T} r^i - m \right ] = 0\quad.&}
\end{eqnarray}
The $S=1$ BW
field function  presents itself a  {\it symmetric} spinor of the second
rank ($4\times 4$ symmetric matrix); the derivative acts to the left hand
in the second equation.  So the field function obeys the Dirac
oscillator equation in each of indices.

The symmetric wave function is expanded in the complete
set of $\gamma$- matrices\footnote{With taking into account
symmetric properties of the field function,
it is  sufficiently to use only $\gamma_\mu C$ and $\sigma_{\mu\nu} C$,
ref~\cite{Lurie} in the considered case;
$C$ is used as the matrix of a charge conjugation. Cf. with the
formula (4), the $S=0$ case, in Ref.~\cite{DV3}.}
\begin{equation}
\Psi_{\{ \alpha\beta \}}= \gamma^\mu_{\alpha\delta} C_{\delta\beta} A_\mu
+ \sigma^{\mu\nu}_{\alpha\delta} C_{\delta\beta} F_{\mu\nu}\quad.
\label{sim}
\end{equation}
The obtained equations for $A_\mu$ and $F_{\mu\nu}$ are the following:
\begin{mathletters} \begin{eqnarray}
\label{P1} \partial_\nu F^{\nu 0} &=&
-{m\over 2} A^0 + {k\over 2} (r^i A^i)\quad,\\ \partial_\nu F^{\nu i} &=&
-{m\over 2} A^i + {k\over 2} r^i A^0\quad,\\ 2m F_{i0} &=& (\partial_i
A_0-\partial_0 A_i) + 2k \, (r^j F_{i}^{\,\,\, j})\quad,\\ \label{P4}
2m F_{jk} &=& (\partial_j A_k-\partial_k A_j) - 2k \, (r^j F^{0}_{\,\,\, k} -
r^k F^{0}_{\,\,\,j})\quad.  \end{eqnarray} \end{mathletters} Let me
introduce $E^i = F^{i0}$ and $B^i =-{1\over 2} \epsilon^{ijk} F^{jk}$ \, .
Then, expressing the
dependence of the wave function on $t$ as $\exp (-i{\cal E}t)$ one can
obtain the equations ($c=\hbar=1$):
\begin{eqnarray} \cases{({\cal E}+m)
E^i + \sqrt{2} \epsilon^{ijk} p_j^{\,+} B^k = {i\over 2} ({\cal E} +m) A^i
-{i\over \sqrt{2}} p_i^{\,+} A^0 &\cr &\cr ({\cal E}-m) E^i + \sqrt{2}
\epsilon^{ijk} p_j^{\,-} B^k = -{i\over 2} ({\cal E} -m) A^i +{i\over
\sqrt{2}} p_i^{\,-} A^0 &\cr &\cr 2 (p_i  E^i) = im A^0 -ik (r^i A^i) &\cr
&\cr
2m\epsilon^{ijk} B^k - 2k (r^i E^j -r^j E^i) = i ( p_i A^j - p_j
A^i)\quad, &\cr}
\end{eqnarray}
where $\vec p^{\, \pm} = {1\over \sqrt{2}}
(\vec p \pm k \vec r)$ and $\vec p_i = {1\over i} \vec\nabla_i$.  This set
can be  re-written in a more symmetric form after the substitution $D^i =
E^i -{i\over 2} A^i$ and $F^i = E^i +{i\over 2} A^i
=(D^i)^\ast$ if $E^i$ and $A^i$ are real quantities. In such a way
one obtains
\begin{eqnarray}\label{eqqu}
\cases{{im\over \sqrt{2}} A^0 = p_i^{\,+} F^i
+ p_i^{\,-} D^i &\cr
&\cr
m B^i = {1\over \sqrt{2}} \epsilon^{ijk}
\left [ p_j^{\,+} F^k -  p_j^{\,-} D^k \right ] &\cr
&\cr
({\cal E}+m) D^i = -\sqrt{2}\epsilon^{ijk} p_j^{\,+} B^k -{i\over
\sqrt{2}}  p_i^{\,+} A^0 &\cr
&\cr
({\cal E}-m) F^i = -\sqrt{2}\epsilon^{ijk} p_j^{\,-} B^k +{i\over
\sqrt{2}} p_i^{\,-} A^0\quad.  &\cr}
\end{eqnarray}
It is possible to eliminate $A^0$ and $B^i$ on using the commutation
relations
$\left [ p_i^+ \, ,\, p_j^- \right ]_- \,=\, ik\delta_{ij}$ and $\left \{
p_i^+ p_j^- - p_j^+ p_i^- \right \} f(\vec r) = k\epsilon^{ijk} \hat L^k
f(\vec r) = ik (\vec S \vec L)_{ij} f (\vec r)$\, . The result is
\begin{mathletters}
\begin{eqnarray}
\label{eqqu1}
m ({\cal E} +m) D^i &=& \left [
-ik (\vec S \vec L)_{ij} -\vec p_k^{\,+} \vec p_k^{\,-} \delta_{ij} \right
] D^j + \left [2(\vec S \vec p^{\,+})^2_{ij} -(\vec p^{\,+})^2\delta_{ij}
\right ] F^j \quad,\\
\label{eqqf1}
m ({\cal E} -m) F^i &=& \left [(\vec
p^{\,-})^2 \delta_{ij}-2(\vec S \vec p^{\,-})^2_{ij} \right ] D^j + \left
[-ik (\vec S \vec L)_{ij} +\vec p_k^{\,-} \vec p_k^{\,+} \delta_{ij}
\right ] F^j \quad,
\end{eqnarray}
\end{mathletters}
$\vec S$ are the spin-1 matrices,   $\vec L$ is the orbital part of
the angular momentum operator.

In order to carry out further decoupling, the
system of equations (\ref{eqqu1}-\ref{eqqf1})
one could try to apply the procedure of ref.~\cite{DV3}.
But, the calculations are more complicated comparing with
the previous work and, moreover, it does not lead us to desirable
result directly. The system is {\it not} decoupled after the first
application of the procedure of ref.~\cite{DV3}. We arrive at
\begin{mathletters}
\begin{eqnarray}
m^2 ({\cal E}^2 - m^2) D^i &=& [ \mbox{dif. op.}_1 ]_{ij} D^j -2ik p_i^+
p_j^+ F^j \,\, ,\\
m^2 ({\cal E}^2 - m^2) F^i &=& [ \mbox{dif. op.}_2 ]_{ij} F^j +2ik p_i^-
p_j^- D^j \,\, ,
\end{eqnarray}
\end{mathletters}
with  complicated operators $\mbox{dif.op}_{1,2}$ on the right-hand
side of the equation. On the other hand, we do not want to apply the
procedure of ref.~\cite[p.298]{Mosh3}, because it is doubtfully that one
can insert the complete set of the state vectors as in the formulas
(108,109) of the cited reference between $\bbox{\eta}\cdot {\bf S}$ and
$\bbox{\xi}\cdot {\bf S}$.  Such bra- vectors as in (108a,108b) may {\it
not} exist, e.~g., in the case of low quantum numbers $N$ and $j$.

Nevertheless, one can use another method. Namely, 1) multiplying,
e.~g., the first equation (\ref{eqqu1}) by $(D^i)^\ast$, 2) integrating
over $d^3 {\bf r}$ and 3) using identities of the hermitian conjugation,
the expansion over spherical tensors and the normalization conditions the
problem is solved. On this basis we derive the quantization rule for the
pure imaginary $k=im\omega$
\begin{equation}
{\cal E} = -{ik \over 2m}
\left [ j(j+1) -l(l+1) -s(s+1) \right ] -{3ik\over 2m}\,\, .
\end{equation}
Very surprisingly, the principal quantum number does {\it not}
present here! The energy is due to
the spin-orbit interaction
and the constant term, which is similar to that appeared in
the Moshinsky-Szczepaniak version of the $S=1/2$ oscillator.
We remain the  interpretation of this fact (as well as of Eqs.
(\ref{nr1},\ref{nr2}) below) for future publications.  Finally, for the
states of parity $(-1)^j$ one has
\begin{equation}
\label{nr1}
{\cal E} = {\hbar
\omega \over 2}\,\, ,
\end{equation}
which can be interpreted
as zero-mode oscillations (we recover $\hbar$ for visual purposes).  On
the other hand, for the states of parity $-(-1)^j$ one has
\begin{equation}
\label{nr2}
{\cal E} = \pm
{\hbar \omega \over 2} ( 2j+1)\,\, ,
\end{equation}
i.~e.  the
{\it non-relativistic} formula for the harmonic oscillator with the
substitution $N \rightarrow j$ and with two signs of the
energy!\footnote{Cf. with the discussion on the page 177 in~\cite{DV3}.}

\section{The Two-Body Relativistic Oscillator}

Now I would like to pay some attention to
the case of the two-body Dirac oscillator~\cite{Mosh2}.
The two-body Dirac Hamiltonian with oscillator-like interaction
is given by ($m \omega=1$ and, hence, $k=i$ here)\footnote{The
oscillatorlike system with $\sim (\vec \alpha_1 -\vec \alpha_2) \cdot \vec
r B\Gamma_5$ has also been considered.}
\begin{eqnarray} \lefteqn{i\left
({\partial \over \partial t_1} + {\partial \over \partial t_2}\right
)\psi={\cal H}\psi=}\nonumber\\ &=&\left [{1\over \sqrt{2}}(\vec
\alpha_1+\vec \alpha_2)\cdot \vec P+{1\over \sqrt{2}}(\vec \alpha_1 -\vec
\alpha_2)\cdot \vec p-{i\over \sqrt{2}}(\vec \alpha_1-\vec \alpha_2)\cdot
\vec r B+m(\beta_1+\beta_2)\right ] \psi,\nonumber\\ &&\label{tbdo}
\end{eqnarray}
where the matrices are given by the direct products
\begin{equation}
\vec \alpha_1=\pmatrix{
0 & \vec\sigma_1 \cr
\vec \sigma_1 & 0 \cr
}\otimes \pmatrix{
\openone_{2\otimes 2} & 0 \cr
0 & \openone_{2\otimes 2} \cr
},\quad \vec \alpha_2=\pmatrix{
\openone_{2\otimes 2} & 0 \cr
0 & \openone_{2\otimes 2} \cr
}\otimes \pmatrix{
0 & \vec \sigma_2 \cr
\vec \sigma_2 & 0 \cr
},
\end{equation}
\begin{equation}
B=\beta_1\otimes \beta_2=\pmatrix{
\openone_{2\otimes 2} & 0 \cr
0 & -\openone_{2\otimes 2} \cr
}\otimes \pmatrix{
\openone_{2\otimes 2} & 0 \cr
0 & -\openone_{2\otimes 2} \cr
},
\end{equation}
\begin{equation}
\Gamma_5 = \gamma_1^5\otimes \gamma_2^5 = \pmatrix{
0 & \openone_{2\otimes 2} \cr
\openone_{2\otimes 2} & 0 \cr
}\otimes \pmatrix{
0 & \openone_{2\otimes 2} \cr
\openone_{2\otimes 2} & 0 \cr
}.
\end{equation}
If we are in the center of mass system (c.m.s.) it is possible
to equate $\vec P=0$. While the two-body Dirac oscillator
seems not to have found considerable  phenomenological applications (cf.
spectra presented in~\cite{Mosh3} with the experiment), this is an
interesting mathematical model.

Now let me apply the same procedure like that which was used for
the transformation of the Bargmann-Wigner set of equations to the Proca
equations (see~\cite{Lurie,DV3} and what is above): the 16- component
wave function of two-body Dirac equation can also be expanded
in the complete set of matrices:
 $(\gamma^\mu C)$, $(\sigma^{\mu\nu} C)$
and $C$, $(\gamma^5 C)$, and $(\gamma^5 \gamma^\mu C)$.
The wave function is decomposed
in symmetric and antisymmetric parts using the above-mentioned
complete system of matrices
\begin{equation}
\psi=\psi_{\{\alpha\beta\} } +\psi_{\left [\alpha\beta\right ]},
\end{equation}
where the first term is given by the formula (\ref{sim}) and the second
term, by the formula (4) of Ref.~\cite{DV3}.
In such a way we obtain the set of equations:\footnote{We corrected here
the misprints  in the signs of the equations of ref.~\cite{DV0}.}
\begin{mathletters}
\begin{eqnarray}
\label{eq1}&&{\cal E} A_0=0, \quad {\cal E}\tilde A_0=-2m\tilde \varphi
\,\, ,\\
\label{eq2}&&{\cal E}\varphi =4i  \vec p_i^{\,-} F^{i0}\,\,,\\
\label{eq3}&&{\cal E}\tilde \varphi = -2m\tilde A_0 +
2\epsilon^{ijk} \vec p_i^{\,+} F^{jk}\,\, ,\\
\label{eq4}&&{\cal E}\tilde A^i = 2i\epsilon^{ijk} \vec p_j^{\,\mp} A^k
\,\, ,\\
\label{eq5}&&{\cal E} A^i = 4im F^{0i} +2i \epsilon^{ijk} \vec p_j^{\,\pm}
\tilde A^k\,\, ,\\
\label{eq6}&&{\cal E} F^{0i} = -2im A^i + 2i \vec p_i^{\,+}
\varphi\,\, ,\\
\label{eq7}&&{\cal E} F^{jk} = \epsilon^{ijk} \vec
p_i^{\,-}\tilde\varphi \,\, .
\end{eqnarray} \end{mathletters}
The signs in the
set (\ref{eq1} - \ref{eq7}) correspond to two types of Dirac
oscillator-like interactions, with $\sim (\vec \alpha_1 - \vec \alpha_2)
B$ and $\sim (\vec \alpha_1 -\vec \alpha_2) B\Gamma_5$, respectively.

The two-body Dirac oscillator equations in the form
(\ref{eq1})-(\ref{eq7}) can be uncoupled into the set containing  only
functions $\varphi$, $\tilde\varphi$ and $\tilde A_\mu$ and the another
one containing only $A_\mu$ and $F_{\mu\nu}$:
\begin{mathletters}
\begin{eqnarray}
&&({\cal E}^2 - 8m^2)\varphi = 8(\vec p_i^{\,-} \vec p_i^{\,+})\varphi
- {16im\over {\cal E}} \,\epsilon^{ijk}\vec p_i^{\,-} p_j^{\,\pm}
\tilde A^{\,k}\,\, ,\\
&&({\cal E}^2 - 4m^2)\tilde \varphi = 4 (\vec p_i^{\,+} \vec p_i^{\,-})
\tilde \varphi\,\, ,\\
&&{\cal E}\tilde A_0 = -2m\tilde
\varphi\,\, ,\\
&& ({\cal E}^2 -8m^2)\tilde A^{\,i} = 4 (\vec p_j^{\,\mp} \vec p_j^{\,\pm})
\tilde A^{\,i} - 4(\vec p_j^{\,\mp} \vec p_i^{\,\pm}) \tilde A^{\,j}
- {16im\over {\cal E}} \,\epsilon^{ijk}
\vec p_{\,j}^{\,\mp}\vec p_k^{\,+} \varphi \,\, ;
\end{eqnarray} \end{mathletters}
and
\begin{mathletters} \begin{eqnarray}
&&{\cal E} A_0 = 0\,\, ,\\
&&{\cal E}^2 A^i = 4im {\cal E} F^{0i} +
4 (\vec p_j^{\,\pm} \vec p_j^{\,\mp}) A^i - 4 (\vec
p_j^{\,\pm} \vec p _i^{\,\mp} ) A^j \,\, ,\\
&&{\cal E}^2 F^{0i} = -2im{\cal E} A^i  -
8 (\vec p_i^{\,+} \vec p_j^{\,-}) F^{0j} \,\, ,\\
&& ({\cal E}^2 - 4m^2) F^{jk} = 2\epsilon^{ijk} \epsilon^{lmn} (\vec
p_i^{\,-} \vec p_l^{\,+}) F^{mn}\,\, .
\end{eqnarray} \end{mathletters}
This fact proves the Dirac oscillator interaction, like the case when we
introduce the (self-) interaction with the transverse 4-vector  potential
into the Proca equation (or, equivalently, into the Bargmann-Wigner
equations), does not mix $S=1$ and $S=0$ states.

{\it Acknowledgments.}
I acknowledge discussions with Profs. D. V. Ahluwalia,
Y. Nedjadi, A. del Sol Mesa and Yu. F. Smirnov.

I am grateful to Zacatecas University for professorship.

\end{document}